\documentclass[letterpaper, 10pt, conference]{ieeeconf}

\IEEEoverridecommandlockouts                              
\usepackage{graphicx}
\usepackage[dvips]{epsfig}

\usepackage{amssymb,amsmath,amsfonts,enumerate}
\usepackage{cite}
\usepackage{caption,subcaption}

\usepackage{csquotes}
\usepackage[UKenglish,USenglish]{babel}

\def\la{\lambda}
\def\La{\Lambda}
\def\vp{\varphi}
\def\ve{\varepsilon}

\def\r{\mathbb R}

\def\be{\beta}

\def\ones{\mathbf 1}

\def\be{\begin{equation}}
\def\ee{\end{equation}}
\def\ben{\begin{equation*}}
\def\een{\end{equation*}}

\newtheorem{thm}{Theorem}
\newtheorem{lem}{Lemma}
\newtheorem{cor}{Corollary}
\newtheorem{defn}{Definition}
\newtheorem{rem}{Remark}

\newtheorem{prop}{Proposition}

\title{New Results on Delay Robustness of Consensus Algorithms}

\author{Anton V. Proskurnikov and Giuseppe Calafiore%
\thanks{The authors are with the Department of Electronics and Telecommunications, Politecnico di Torino, Turin, Italy.
Anton V. Proskurnikov is also with Institute for Problems of Mechanical Engineering, Russian Academy of Sciences,
St. Petersburg, Russia. 
Email: {\tt\small anton.p.1982@ieee.org, giuseppe.calafiore@polito.it}
}%
}

\begin{document}

\maketitle
\thispagestyle{empty}
\pagestyle{empty}

\begin{abstract}
Consensus of autonomous agents is a benchmark problem in cooperative control. In this paper, we consider standard continuous-time averaging consensus policies (or Laplacian flows) over time-varying graphs and focus on robustness of consensus against \emph{communication delays}. Such a robustness has been proved under the assumption of uniform quasi-strong connectivity of the graph. It is known, however, that the uniform connectivity is not necessary for consensus. For instance, in the case of undirected graph and undelayed communication consensus requires a much weaker condition of integral connectivity. In this paper, we show that the latter results remain valid in presence of unknown but bounded communication delays, furthermore, the condition of undirected graph can be substantially relaxed and replaced by the conditions of non-instantaneous type-symmetry. Furthermore, consensus can be proved for any feasible solution of the delay differential inequalities associated to the consensus algorithm. Such inequalities naturally arise in problems of containment control, distributed optimization and models of social dynamics.
\end{abstract}

\section{Introduction}

Consensus policies are prototypic distributed algorithms for multi-agent coordination~\cite{RenBeardBook,RenCaoBook} inspired by ``regular'' intelligent behaviors of biological, physical and social systems~\cite{Vicsek,Strogatz:00,Couzin:2011,ProTempo:2017-1}.
The most studied first-order consensus algorithms are based on the principle of iterative averaging; such algorithms were studied in the literature~\cite{French:1956,Abelson:1964,DeGroot,Tsitsiklis:86} long before the recent ``boom'' in multi-agent control and network science. Averaging policies have found numerous applications, being, in particular, an indispensable part of distributed algorithms for optimization and equation solving~\cite{ShiJohanssonHong:13,Nedic:10,Wang2019_ARC,FullmerMorse2018}.

In this paper, we deal with continuous time averaging dynamics, or \emph{Laplacian flows} over graphs~\cite{BulloBook-Online}. Consider a team of agents indexed $1$ through $n$ and associated with some values of interest $x_i\in\r$, which can stand e.g. for attitudes~\cite{Abelson:1964} or some physical characteristics~\cite{Murray:07}. In the simplest situation, the values are governed by the equations
\be\label{eq.conse1}
\dot x_i(t)=\sum_{j=1}^n a_{ij}(t)(x_j(t)-x_i(t))\quad\forall i=1,\ldots,n,
\ee
where $A(t)=(a_{ij}(t))$ is a nonnegative matrix of \emph{influence weights}. In other words, the value of agent $i$ is attracted by the values of ``adjacent'' agents $j$ with $a_{ij}(t)>0$, and the weight $a_{ij}$ characterizes the strength of such an attraction.

The central question regarding dynamics~\eqref{eq.conse1} is establishing eventual \emph{consensus}, that is, convergence of all values $x_i(t)$ to the same value $\bar x=\lim_{t\to\infty}x_i(t)\,\forall i$ (obviously, $\bar x$ depends on the initial condition).  More general behavior is ``partial'' consensus (group consensus, clustering)~\cite{Yu10,XiaCao:11}, that is, splitting of the agents into several groups that agree on different values. In the case of constant weights $A(t)\equiv A$, the consensus criterion is well-known~\cite{ChebotarevAgaev:2002,RenBeardBook} and reduces to the quasi-strong connectivity of the graph associated to matrix $A$ (equivalently, the graph has a directed spanning tree). Informally, some agent influences all other agents in the network, directly or indirectly. Without this ``weak'' connectivity, the values $x_i(t)$ converge, and their steady values are determined by the graph's spanning forest~\cite{AgaevChe:2014}.

To find criteria ensuring consensus in the case of a general time-varying matrix $A(t)$ is a difficult problem whose complete solution is still elusive. A well-known \emph{necessary} condition for consensus is the so-called \emph{integral} (essential, persistent) connectivity~\cite{TsiTsi:13,MatvPro:2013,ShiJohansson:13-1}. Namely, the ``persistent'' arcs corresponding to such pairs of agents that
\be\label{eq.int-aij}
\int_0^{\infty}a_{ij}(t)dt=\infty
\ee
should constitute a quasi-strongly connected graph. This condition, however, is far from being sufficient: a simple counter-examples with $n=3$ agents in~\cite{Moro:05} shows that the agents may fail to reach consensus (and, moreover, their values do not converge) even if the graph of persistent interactions is complete. This effect is caused by imbalance between the different couplings: some of them are stronger than others (the integrals in~\eqref{eq.int-aij} diverge at different rates).

Sufficient criteria for consensus can be divided into two groups. Conditions of the first type require the uniform quasi-strong connectivity~\cite{Moro:04,RenBeardBook,LinFrancis:07}.
Omitting some technical details, the uniform quasi-strong connectivity implies that for some time horizon $T$ the unions of the interacting graphs over each interval $[t,t+T]$ is quasi-strongly connected. This condition is not necessary for consensus in the usual sense and implies, in fact,
much stronger properties of consensus with exponential convergence~\cite{BarabanovOrtega2018} and robust consensus in presence of unknown but bounded noises~\cite{ShiJohansson:13}. Also, the uniform quasi-strong connectivity implies consensus in more general nonlinear averaging schemes~\cite{LinFrancis:07,Antonis,Muenz:11}.

Consensus criteria of the second kind ensure consensus in presence of the integral connectivity and some conditions preventing the imbalance of couplings. The simplest condition of this type is the coupling symmetry $a_{ij}(t)=a_{ji}(t)$ (the interaction graph is undirected)~\cite{CaoZheng:11}.
The latter condition may be in fact replaced by the weight-balance, type-symmetry and cut-balance conditions~\cite{TsiTsi:13} (the relevant definitions are given below). All of these conditions guarantee \emph{reciprocity} of interactions: if some group of agents $S\subset\{1,\ldots,n\}$ influences the remaining agents from $S^c=\{1,\ldots,n\}\setminus S$, then agents $S^c$ also influence agents from $S$, moreover, these mutual influences are commensurate. The most recent results from~\cite{MartinHendrickx:2016} allow non-instantaneous forms of the latter property, for instance, the action of group $S$ on group $S^c$ may be responded after some limited amount of time. It is remarkable that some conditions of reciprocity imply consensus not only in the ODE system~\eqref{eq.conse1}, but also in the system of associated \emph{differential inequalities}~\cite{ProCao:2017}
\be\label{eq.ineq1}
\dot x_i(t)\leq\sum_{j=1}^n a_{ij}(t)(x_j(t)-x_i(t))\quad\forall i=1,\ldots,n.
\ee
As discussed in~\cite{ProCao:2017}, the inequalities of this type naturally arise in analysis of many cooperative control algorithms, e.g. containment control, target surrounding and distributed optimization, and some models of opinion dynamics.

An important question regarding consensus algorithms is robustness against \emph{communication delays}. Such delays naturally arise in the situation where the agents have direct access to their own values, whereas the neighbors' values are subject to non-negligible time lags. Delays of this type are inevitable in networks spread over large distances (where the agents e.g. communicate via Internet), but also arise in many physical models~\cite{StrogatzDelay1:03,Antonis,DahmsSchoell:2012}. It is well-known that consensus criterion of the first type (uniform connectivity) guarantee consensus robustness against arbitrary bounded delays~\cite{Moro:04,Antonis,Muenz:11}. However, delay robustness of consensus ensured by the conditions of the second kind (integral connectivity and reciprocity) remains an open problem, since all the relevant works~\cite{TsiTsi:13,MatvPro:2013,MartinHendrickx:2016,MartinGirard:2013} are confined to systems of ordinary differential equations~\eqref{eq.conse1}.

In this paper, we extend the existing results on delay robustness in two ways. We show that for a general directed graph, the condition of uniform connectivity can be replaced by a condition that we call \emph{repeated} connectivity. This relaxation allows, for instance, arbitrarily long periods of ``silence'' where the agents do not communicate (intermittent communication graphs~\cite{Guanghui:12}). In the case where the matrix of weights is non-instantaneously type-symmetric~\cite{MartinHendrickx:2016}, the repeated connectivity can be further relaxed to the standard condition of integral connectivity. Finally, we show that the criteria of consensus derived in this paper remain valid for the delayed modification of differential inequality~\eqref{eq.ineq1}.

\section{Preliminaries}

Henceforth, we denote $\ones_n=(1,1,\ldots,1)^{\top}\in\mathbb{R}^n$. The standard coordinate basis vectors are denoted by $\mathbf{e}_1=(1,0,\ldots,0)^{\top},\ldots,\mathbf{e}_n=(0,0,\ldots,1)^{\top}$. The symbol $I_n$ denotes $n\times n$ identity matrix. For two vectors $x,y\in\r^n$, we write $x\leq y$ if $x_i\leq y_i\,\forall i$. The set $\{m,m+1,\ldots,n\}$, where $m\leq n$, is denoted by $[m:n]$.

\subsection{Matrices and graphs}

To every non-negative square matrix $A=(a_{ij})_{i,j\in V}$ we associate the directed weighted graph $G[A]=(V,E[A],A)$, whose nodes constitute the finite set $V$ and whose arcs correspond to non-zero entries\footnote{Following the convention adopted in multi-agent systems~\cite{RenBeardBook}, influence of agent $j$ on agent $i$ is represented by arc $(j,i)$ rather than $(i,j)$.}
$E[A]=\{(j,i):a_{ij}>0\}$. The weighted graph is called \emph{undirected} if its matrix is symmetric $A=A^{\top}$.

Given a general directed graph $G=(V,E)$, where $E\subseteq V\times V$, the walk from node $i\in V$ to node $j\in V$ is a sequence of arcs $(v_0,v_1)$, $(v_1,v_2)$,\ldots,$(v_{n-1},v_n)$ starting at $i_0=i$ and ending at $i_n=j$. A graph is \emph{strongly connected} if every two nodes are connected by a walk and \emph{quasi-strongly connected} (or has a directed spanning tree) if some node (a \emph{root}) is connected to all other nodes by walks.

A time-varying matrix $A(t)=(a_{ij}(t))$ is called \emph{type-symmetric} if  there exists a constant $K\geq 1$ such that
\be\label{eq.type-symm}
K^{-1}a_{ji}(t)\leq a_{ij}(t)\leq Ka_{ji}(t)\quad\forall i,j\in V\,\forall t\geq 0.
\ee
Obviously, symmetric matrix is type-symmetric (with $K=1$).
The type-symmetry condition requires bidirectional communication between the agents: $a_{ij}(t)>0$ if and only if $a_{ji}(t)>0$. A natural relaxation of this property is the \emph{non-instantaneous}  type-symmetry introduced in~\cite{MartinHendrickx:2016}.
\begin{defn}\label{def.symm}
The matrix function $A(\cdot)$ is said to satisfy the non-instantaneous type-symmetry property if
there exist an increasing sequence $0=t_0<t_1<\ldots$, $t_p\to\infty$ and a constants $K\geq 1$ such that for any $i,j=1,\ldots,n$ one has
\begin{gather}\label{eq.type-symm-non}
K^{-1}\int\limits_{t_p}^{t_{p+1}}a_{ji}(t)\,dt\leq \int\limits_{t_p}^{t_{p+1}}a_{ij}(t)\,dt\leq K\int\limits_{t_p}^{t_{p+1}}a_{ji}(t)\,dt.
\end{gather}
\end{defn}

Verification of~\eqref{eq.type-symm-non} may seem quite non-trivial, because the sequence $t_p$ should be same for all pairs $(i,j)$. In reality, this condition can be efficiently tested, as follows from the proof of Theorem~2 in~\cite{MartinHendrickx:2016}. The work~\cite{MartinHendrickx:2016} contains also some practical examples of directed networks, obeying~\eqref{eq.type-symm-non}.

Along with the non-instantaneous type-symmetry, we will consider another condition ensuring (under some technical assumptions) consensus, called \emph{repeated strong connectivity}.
\begin{defn}\label{def.repeated}
We call the matrix function $A(\cdot)$ \emph{repeatedly strongly connected} if there exists a sequence $0=t_0<t_1<\ldots<t_p<\ldots$, $t_p\to\infty$ and a constant $\ve>0$ such that the graphs $G_{p,\ve}=(V,E_{p,\ve})$ are strongly connected, where
\be\label{eq.g-pe}
E_{p,\ve}=\left\{(j,i):\int_{t_p}^{t_{p+1}}a_{ij}(t)\geq\ve\right\}.
\ee
\end{defn}

In other words, the unions of graphs over intervals $[t_p,t_{p+1}]$ are strongly connected and retain this property after removing ``lightweight'' arcs. The special case of condition from Definition~\ref{def.repeated} with $t_p=pT$, where $T$ is some constant period, is known as the \emph{uniform strong connectivity}~\cite{ProCao:2017}. A simple example where the matrix is repeatedly yet non-uniformly strongly connected is the intermittent communication graph~\cite{Guanghui:12}, where $A(t)$ switches between some constant matrix $A_0$ (with strongly connected graph) and the matrix of zeros. The periods of silence can be arbitrarily long, which destroys the uniform connectivity, however, repeated connectivity can be easily proved.

Finally, given a nonnegative matrix $A=(a_{ij})_{i,j\in V}$, we introduce \emph{persistent graph} $G_{\infty}=(V,E_{\infty})$, where $E_{\infty}$ consists of all pairs $(j,i)$ such that condition~\eqref{eq.int-aij} holds.

\subsection{Linear delay systems}

In the next sections, we deal with linear delay systems\footnote{To simplify notation, we consider only discrete (lumped) delays, although the proofs also work for distributed delays introduced in~\cite{Muenz2009}.}
\be\label{eq.delay}
\dot x(t)=\sum_{i=0}^sP_i(t)x(t-h_i(t))+f(t)\in\r^n, t\geq t_0,
\ee
where $P_1,\ldots,P_s$ are some $n\times n$ matrices and $h_i(t)\in [0,\bar h]$ are time-varying uniformly bounded delays. We will also consider systems of associated differential inequalities
\be\label{eq.delay1}
\dot y(t)\leq\sum_{i=0}^sP_i(t)y(t-h_i(t))+f(t), t\geq t_0.
\ee
By default, a solution of~\eqref{eq.delay} (respectively,~\eqref{eq.delay1}) is a vector function $x(t)\in\r^n$ (respectively, $y(t)\in\r^n$) that is defined and \emph{locally bounded}\footnote{A function is locally bounded if it is bounded on every compact set; this property holds e.g. for continuous functions.} on $[t_0-\bar h,\infty)$, \emph{absolutely continuous} on $[t_0,\infty)$ and satisfies~\eqref{eq.delay} (respectively,~\eqref{eq.delay1}) for almost all $t\geq t_0$.
Without loss of generality, we may assume that the solution is \emph{right-continuous} at $t=t_0$, the left continuity at $t=t_0$ and continuity at $t<t_0$ are however not required.

For a function $x:[t_0-\bar h,\infty)\to\r^n$ and $t\ge t_0$, let
\[
x^t:[-\bar h,0]\to\r^n,\quad x^t(\theta)=x(t+\theta)\,\forall\theta\in [-\bar h,0].
\]
The general criteria of solutions existence and uniqueness~\cite{Hale} imply that for locally bounded functions $P_i(t),f(t)$ of appropriate dimensions, any initial condition
\be\label{eq.initial}
x(t_0)=a\in\r^n,\quad x^{t_0}=\vp\in L_{\infty}([-\bar h,0]\to\r^n)
\ee
determines the unique solution $x(t)=x(t|t_0,a,\vp,f)$ of~\eqref{eq.delay}.
Similar to the undelayed case, the Cauchy formula~\cite{Hale}
\be\label{eq.cauchy}
\begin{aligned}
&x(t|t_0,a,\vp,f)=x(t|t_0,a,\vp,0)+\int_{t_0}^tU(t,\xi)f(\xi)d\xi\\
&=x(t|t_0,0,\vp,0)+U(t,t_0)a+\int_{t_0}^tU(t,\xi)f(\xi)d\xi.
\end{aligned}
\ee
can be derived. Here $U(t,s)$ is the $n\times n$ \emph{evolutionary  matrix}, defined as the unique solution to the Cauchy problem
\be\label{eq.u}
\begin{gathered}
\frac{\partial}{\partial t} U(t,\xi)=\sum_{i=0}^sP_i(t)U(t-h_i(t)),\quad t\geq \xi,\\
U(\xi,\xi)=I_n,\quad U(t,\xi)\equiv 0\,\forall t<\xi.
\end{gathered}
\ee

\begin{rem}\label{rem.positive}
If the evolutionary matrix $U(t,s)$ is \emph{non-negative}, then a counterpart of the ``comparison lemma'' holds: for any solution of the inequality~\eqref{eq.delay1} with initial condition~\eqref{eq.initial} one has $y(t)\leq x(t|t_0,a,\vp,f)\,\forall t\geq t_0$.

Indeed,~\eqref{eq.delay1} implies that $y(t)=x(t|t_0,a,\vp,f-g)$, where $g(t)=\sum_{i=0}^sP_i(t)y(t-h_i(t))+f(t)-\dot y(t)\geq 0$. The Cauchy formula~\eqref{eq.cauchy} entails that $x(t|t_0,a,\vp,f-g)\leq x(t|t_0,a,\vp,f)$.
\end{rem}

\section{Main results}

In this section, we present our main result, establishing convergence properties for the generalization of system~\eqref{eq.conse1}
\be\label{eq.conse2}
\begin{gathered}
\dot x_i(t)=\sum_{j=1}^na_{ij}(t)(\hat x_j^i(t)-x_i(t)),\,i\in [1:n],\\
\hat x_j^i(t)=x_j(t-{h}_{ij}(t)).
\end{gathered}
\ee
Along with the delay equations~\eqref{eq.conse2}, we consider an associated system of delay differential inequalities
\be\label{eq.ineq2}
\begin{gathered}
\dot x_i(t)\leq\sum_{j=1}^na_{ij}(t)(\hat x_j^i(t)-x_i(t)),\,i\in [1:n].
\end{gathered}
\ee
As discussed in~\cite{ProCao:2017}, convergence of many algorithms of multi-agent coordination, based on Laplacian flow dynamics, reduces to analysis of inequalities~\eqref{eq.ineq1}, namely, convergence of all (bounded) solutions and behaviors of the \emph{residuals}
\be\label{eq.residual}
\Delta_i(t)=\sum_{j=1}^na_{ij}(t)(\hat x_j^i(t)-x_i(t))-\dot x_i(t), i\in [1:n].
\ee
The examples considered in~\cite{ProCao:2017} can be extended to the delayed case, using the results of Theorem~\ref{thm.1} presented below.

Equation~\eqref{eq.conse2} and inequality~\eqref{eq.ineq2} are determined by the nonnegative matrices $(a_{ij}(t))$ and $(h_{ij}(t))$. The entry $h_{ij}(t)$ stands for the time lag in the value from agent $j$, received by agent $i$ at time $t$. By definition, $h_{ii}(t)\equiv 0$, the other delays are uniformly bounded $0\leq h_{ij}(t)\leq \bar h<\infty$. To simplify matters, we also suppose that the weights are bounded
\be\label{eq.bound-weight}
a_{ij}(t)\leq \bar a\quad\forall t\ge 0.
\ee
Without loss of generality, we also assume that $a_{ii}(t)\equiv 0$.

Obviously, every solution of~\eqref{eq.conse2} is a feasible solution of inequality~\eqref{eq.ineq2}. The inequality, however, has many other solutions, some of them can be unbounded. It will be shown that all feasible solutions of~\eqref{eq.ineq2} have a finite \emph{upper} bound yet may be unbounded from below (e.g. converge to $-\infty$).

We are now ready to formulate our main result.

\begin{thm}\label{thm.1}
Let one of the following conditions hold:
\begin{enumerate}[i)]
\item the matrix $A(\cdot)$ is non-instantaneously type-symmetric~\eqref{eq.type-symm-non} and the graph $G_{\infty}$ is connected\footnote{It follows from~\eqref{eq.type-symm-non} that the graph $G_{\infty}$ is undirected: if $a_{ij}\not\in L_1[0,\infty)$, then also $a_{ji}\not\in L_1[0,\infty)$ and vice versa.};
\item the matrix $A(\cdot)$ is repeatedly strongly connected.
\end{enumerate}
Assume also that the corresponding sequence $t_p$ (from Definition~\ref{def.symm} or~\ref{def.repeated}) satisfies additional constraint:
\be\label{eq.bound}
M:=\max_{i,j}\sup_{p\geq 0}\int_{t_p}^{t_{p+1}}a_{ij}(t)dt<\infty.
\ee
Then, every solution of the inequality~\eqref{eq.ineq2} exhibits eventual consensus in the sense that the limit exists
\be\label{eq.x-bar}
x^*=\lim_{t\to\infty} x(t)
\ee
whose components are coincident $x_1^*=\ldots=x_n^*=c^*\geq -\infty$. If $c^*>-\infty$ (the solution remains bounded), then also
\be\label{eq.delta0}
\int_{t}^{t+T}\Delta(\xi)d\xi\xrightarrow[t\to\infty]{}0.
\ee
for every horizon $T>0$.
\end{thm}
\begin{rem}
In the case where $c_*=-\infty$, Theorem~\ref{thm.1} does not guarantee that $x_i-x_j\xrightarrow[t\to\infty]{}0.$
Hence, for unbounded solutions consensus is a rather weak property that does not imply the asymptotic synchrony of all components $x_i(t)$.
\end{rem}
\begin{rem}
Obviously, every solution to~\eqref{eq.conse2} is a feasible solution to~\eqref{eq.ineq2}, which is automatically bounded (see Proposition~\ref{prop.bound} below). Hence, Theorem~\ref{thm.1} guarantees consensus in the
usual consensus algorithm~\eqref{eq.conse2}. For the equations~\eqref{eq.conse2}, the condition of repeated strong connectivity can be relaxed to repeated quasi-strong connectivity (the graphs $G_{p,\ve}$ are quasi-strongly connected, or have a directed spanning tree). 

Consensus in inequalities, however, requires strong connectivity even in the undelayed case~\cite{ProCao:2017}. Notice that consensus under \emph{repeated} strong connectivity does not follow from previously known results~\cite{Antonis,Muenz:11} since, first, we do not require the weights $w_{ij}(t)$ to be piecewise-continuous and, second, the repeated strong connectivity does not follow from uniform connectivity commonly adopted in the literature.
\end{rem}
\begin{rem}
Assumption~\eqref{eq.bound} cannot be discarded even if $\bar h=0$. The counterexample constructed in~\cite[Section~III-B]{MartinHendrickx:2016} deals with a system~\eqref{eq.conse1} with $n=3$ agents. Both non-instantaneous type-symmetry and repeated strong connectivity conditions hold, however, some solutions do not converge.
\end{rem}


It should be noticed that in the case of usual type-symmetry~\eqref{eq.type-symm} and undelayed protocol~\eqref{eq.conse1} stronger convergence properties can be proved~\cite{TsiTsi:13,MatvPro:2013}, namely, the functions $a_{ij}(x_j-x_i)$ and $\dot x_i$ belong to $L_1[0,\infty)$ for all $i,j$. These properties are not guaranteed by Theorem~\ref{thm.1}, and their validity remains an open problem.

\section{Technical propositions}

In this section, we summarize some technical properties of delayed consensus protocol~\eqref{eq.conse2} and inequality~\eqref{eq.ineq2} to be used in the proofs of main results.
The proofs of these technical propositions will be given in Section~\ref{sec.proof}.

Throughout this section, $\alpha_i(t)$ stands for the weighted degree of node $i$, that is,
\be\label{eq.alpha}
\alpha_i(t):=\sum_{j=1}^n a_{ij}(t)\in [0,(n-1)\bar a].
\ee
Using~\eqref{eq.bound}, one may easily notice that
\be\label{eq.bound-alfa}
\begin{gathered}
\int_{\xi}^{\eta}\alpha_i(t)dt\leq \bar\alpha_k:=(n-1)M(k+2)\quad\forall i\\
\forall \xi\in (t_{p-1},t_p],\;\forall\eta\in [t_{p+k},t_{p+k+1}).
\end{gathered}
\ee

For any solution of~\eqref{eq.conse2}, we introduce the maximal and minimal values over the time window $[t-\bar h,t]$ as follows
\be\label{eq.la}
\la(t)=\min_i\inf\limits_{[t-\bar h,t]}x_i(t),\,\La(t)=\max_i\sup\limits_{[t-\bar h,t]}x_i(t).
\ee

\begin{prop}\label{prop.bound}
The function $\La(t)$ is \emph{non-increasing} for any feasible solution of~\eqref{eq.ineq2}.
The function $\la(t)$ is \emph{non-decreasing} for any solution of~\eqref{eq.conse2}.
\end{prop}

Notice that the proofs of Proposition~\ref{prop.bound} given in~\cite{Antonis,Muenz:11} (for the equations~\eqref{eq.conse2}) are in applicable to our situation, since we do not assume the matrix $A(t)$ to be piecewise-constant.
Proposition~\ref{prop.bound} implies the following important property of the Cauchy evolutionary matrices.

\begin{cor}\label{cor.cauchy}
For any nonnegative locally summable matrix $A(\cdot)$ the evolutionary matrix $U(t,\xi)$ of~\eqref{eq.delay1} is \emph{substochastic}, that is, $U(t,\xi)$ is nonnegative and $U(t,\xi)\ones_n\leq \ones_n$. Furthermore, a constant $\psi$ exists such that $U(t,\xi)\ones_n\geq \psi\ones_n$.
\end{cor}

Notice that in the case undelayed case ($\bar h=0$), the Cauchy evolutionary matrix is known to be stochastic~\cite{RenBeardBook}.

The proof of Theorem~\ref{thm.1} in the case of type-symmetric matrix is based on the following lemma.
\begin{lem}\label{lem.tech}
If the matrix $A(\cdot)$ is non-instantaneously type-symmetric~\eqref{eq.type-symm-non}, then
the evolutionary matrices $U(t,\xi)$ have uniformly positive row sums. Formally,
a constant $\varrho=\varrho(n,M,K)>0$ exists such that
\be\label{eq.column}
U(t,\xi)^{\top}\ones_n\geq \varrho\ones_n\quad\forall \xi\geq 0\,\forall t\ge \xi.
\ee
\end{lem}


\section{Proofs}\label{sec.proof}

We start with proving technical propositions.

\subsection{Proof of Proposition~\ref{prop.bound}}

We prove only the first statement of Proposition~\ref{prop.bound}, the second statement reduces to it by considering the solution $(-x(t))$. Consider a solution of~\eqref{eq.ineq2}. Let $\xi\geq 0$ and $\La'>\La(\xi)$. We are going to show that $x_i(t)<\La'\,\forall i$ for any $t\geq\xi$. Obviously, the latter inequalities holds when $t$ is close to $\xi$; let $t_*$ be the \emph{first} instant $t>\xi$ when one of them is violated. Hence,
\[
\begin{gathered}
x_i(t)<\La'\quad\forall i\in[1:n],\,\forall t\in [\xi-\bar h,t_*),\\
x_{j}(t_*)=\La'\quad\text{for some $j$.}
\end{gathered}
\]
However, this contradicts to~\eqref{eq.ineq2} since
\be\label{eq.aux0}
\begin{gathered}
\dot x_j(t)\leq \alpha_j(t)[\La'-x_j(t)]\,\forall t\in [\xi,t_*),\\
x_j(t_*)\leq e^{-\int_{\xi}^{t_*}\alpha_j(s)ds}x_j(\xi)+\La'\left(1-e^{-\int_{\xi}^{t_*}\alpha(s)ds}\right)<\La'.
\end{gathered}
\ee
The contradiction proves that $\La(t)<\La'\;\;\forall t\ge \xi$. Since $\La'>\La(\xi)$ can be arbitrary, we have $\La(t)\leq \La(\xi)$ whenever $t\geq \xi$, i.e., $\La$ is a non-increasing function. $\blacksquare$

\subsection{Proof of Corollary~\ref{cor.cauchy}}

Notice that the function $x(t,\xi,v):=U(t,\xi)v$ is nothing else than the solution of~\eqref{eq.conse2} with initial conditions
\be\label{eq.initial1}
x(\xi)=v,\quad x^{\xi}(\theta)\equiv 0.
\ee
For any vector $v\geq 0$, the solution $x(t)=x(t,\xi,v)$ is nonnegative in view of Proposition~\ref{prop.bound}. Therefore, $U(t,\xi)$ is a nonnegative matrix. If $v=\ones_n$, Proposition~\ref{prop.bound} entails that $x(t)=U(t,\xi)v\leq\ones_n$, and hence $U(t,\xi)$ is substochastic. Also, for $v=\ones_n$ one has
\be\label{eq.aux1}
\dot x_i(t)\geq -\alpha_i(t)x_i(t)\quad\forall t\ge\xi.
\ee
Therefore, for $t\in [\xi,\xi+h]$ we have $x_i(t)\geq\psi_0:=e^{-(n-1)\bar a\bar h}\,\forall i$. In view of Proposition~\ref{prop.bound}, we have $x_i(t)\geq\psi$ for all $t\geq\xi$, i.e., $x(t)=U(t,\xi)\ones_n\geq\psi\ones_n$.
$\blacksquare$

\subsection{Proof of Lemma~\ref{lem.tech}}

We are going to prove the statement of Lemma~\ref{lem.tech} via induction on $n=1,2,\ldots$. The key observation is that any \emph{submatrix} $\tilde A=(a_{ij})_{i,j\in I}$ of matrix $A$ that is non-instantaneously type-symmetric~\eqref{eq.type-symm-non} also satisfies~\eqref{eq.type-symm-non}.

The induction base $n=1$ is obvious. In this situation, the unique agent's dynamics is trivial $\dot x_1=0$ and $U(t,\xi)=1$.

Suppose that~\eqref{eq.column} has been proved for $\leq n-1$ agents, our goal is to prove for $n$ agents. Notice that~\eqref{eq.column} can be reformulated as follows:

\emph{(A) Assume that $\xi\geq 0$. If the vector $v\geq 0$ has at least one component $\geq 1$, then a constant $\gamma=\gamma(v)$ exists such that $\ones_n^{\top}x(t,\xi,v)\geq\gamma,\forall t\geq\xi$.}

Indeed, if (A) holds, then~\eqref{eq.column} holds with $\varrho=\min\gamma(\mathbf{e}_i)$. The inverse statement is straightforward ($\gamma(v)=\varrho\ones_n^{\top}v$).

Statement (A) will be proved via \emph{backward induction} on the number $k=n,n-1,\ldots,1$ of components $i$ such that $v_i\geq 1$. The induction base $k=n$ follows from the second statement of Lemma~\ref{lem.tech} ($\gamma=\psi$).

Assume that for $k=r+1,\ldots,n$ the statement (A) is proved. We now have to prove it for the vector $v\geq 0$ such that at least $r$ of its components are $\geq 1$.

Without loss of generality,
we may assume that $v$ is binary, that is, $v_i=1$ or $v_i=0$. Indeed, if the statement is proved for such vectors, then we may consider the vector
$\tilde v$, where $\tilde v_i=1$ if $v_i\geq 1$ and $\tilde v_i=0$ otherwise. By assumption, $v\geq\tilde v$ and, since $U(t,\xi)$ is a nonnegative matrix, $x(t,\xi,v)\geq x(t,\xi,\tilde v)$.
Renumbering the components, we may suppose that $v_1=\ldots=v_r=1$ (without loss of generality) and $v_i=0$ for $i>r$. In view of Proposition~\eqref{prop.bound}, the agent values $x_i(t)=x_i(t,\xi,v)$ stay in $[0,1]$ for $t\ge\xi$. Also, we may assume without loss of generality that $\xi=t_p$ for some $p$. Indeed, if the induction step is proved for $t_p$ and $t_{p-1}<\xi<t_p$, then $x_i(t)\geq c>0$ for all $i\leq r$ and $t\in [\xi,t_p]$ due to~\eqref{eq.aux1} and~\eqref{eq.bound-alfa}, where $c$ is some constant. Applying (A) for $v=c^{-1}x(t_p)$, $\xi=t_p$ and using the system's linearity, one shows that  $\ones_n^{\top}x(t)$ is uniformly positive for any $t\geq\xi$.

Consider now two delay system of dimensions $r$ and $n-r$:
\begin{gather}
\dot x_i=\sum\limits_{j\leq r}a_{ij}(t)(\hat x_j^i(t)-x_i(t))+f_i(t),\,i=1,\ldots,r;\label{eq.subsys1}\\
\dot x_i=\sum\limits_{j> r}a_{ij}(t)(\hat x_j^i(t)-x_i(t))+f_i(t),\,r<i\leq n.\label{eq.subsys2}
\end{gather}
(here, similar to~\eqref{eq.conse2}, $\hat x^i_j(t)=x_j(t-h_{ij}(t))$). Let $U^+,U^{\dagger}$ stand for the evolutionary matrices of systems~\eqref{eq.subsys1} and~\eqref{eq.subsys2} respectively.
The proof of Corollary~1 shows that $U^+\ones_r\geq\psi\ones_r$, where the constant $\psi$ is same as for the original matrix $U$. According to the hypothesis of forward induction (on $n$), the matrix $U^{\dagger}$ satisfies~\eqref{eq.column} with some constant $\varrho^{\dagger}>0$.

If $x_i(t)>\psi/2$ for all $i\geq r$, $t\geq \xi$, then our statement is proved ($\gamma=r\psi/2$).
Otherwise, let first instant such that $x_i(t_*)=\psi/2$ for some $i\leq r$.
We introduce two constants
\be\label{eq.const-aux}
\begin{gathered}
\beta_1=\frac{\psi e^{-(n-1)M}}{2},\;\beta_2=\frac{\varrho^{\dagger}C\beta_1}{\varrho^{\dagger}C+n-r},\\
C:=\frac{\psi}{2Kr(n-r)},
\end{gathered}
\ee
where $K\geq 1$ is the constant from~\eqref{eq.type-symm-non}. Notice that if
$t_{s-1}<t_*\leq t_s$, where $s\geq p$, then, in view of~\eqref{eq.aux1} and~\eqref{eq.bound-alfa}, we have $x_i(t)\geq\beta_1$ for all $i\leq r$ and $t\leq t_s$.
We are going to show that an index $j>r$ and an instant $t\leq t_s$ exist such that $x_j(t)\geq\beta_2$. Assume, one contrary, that $x_j(t)<\beta_2$ on $[\xi,t_*]$ for all $j<r$.

Obviously, $x(t)=x(t,\xi,v)$ is a solution to~\eqref{eq.conse2}, then the subvectors
$x^+(t)=(x_i(t))_{i\leq r}$, $x^{\dagger}(t)=(x_i(t))_{i>r}$ are solutions, respectively, to~\eqref{eq.subsys1} and~\eqref{eq.subsys2}, where
\be\label{eq.aux-f}
\begin{gathered}
f_i(t)=
\begin{cases}
\sum_{j>r}a_{ij}(t)[\hat x^i_j(t)-x_i(t)]\leq 0\quad\forall i\leq r\\
\sum_{j\leq r}a_{ij}(t)[\hat x^i_j(t)-x_i(t)]\geq 0\quad\forall i>r.
\end{cases}
\end{gathered}
\ee
According to our assumptions, $0\leq\hat x^i_j(t)\leq \beta_2\leq x_i(t)\leq 1$ when $i\leq r<j$ and
$1\geq\hat x^i_j(t)\geq \beta_1\geq x_i(t)\geq 0$ if $i>r\geq j$ (here $t\in [\xi,t_s]$).
 Using the Cauchy formula~\eqref{eq.cauchy} (where $t_0$ is replaced by $\xi=t_p$), one shows that
\[
x^+(t_*)=U^+(t_*,t_p)\ones_r-\int_{t_p}^{t_*}U^+(t_*,t)f^+(t)dt,
\]
and hence $x_i(t_*)\geq \psi-\sum_{l\leq r}\int_{t_p}^{t_s}f_l(t)dt\,\forall i\leq r$. By definition, for some $i\leq r$ we have $x_i(t_*)\leq\psi/2$, and thus for some pairs of indices $l\leq r,j>r$ one has
\be\label{eq.auxaux}
\int_{t_p}^{t_*}a_{lj}(t)[\hat x^l_j(t)-x_l(t)]dt\geq c:=\frac{\psi}{2r(n-r)}.
\ee
Recalling that $0\leq x^l_j(t)-x_l(t)\leq 1$, one shows that
\[
\int_{t_p}^{t_s}a_{lj}(t)dt\geq\int_{t_p}^{t_*}a_{lj}(t)dt\geq c\overset{\eqref{eq.type-symm-non}}{\Longrightarrow}
\int_{t_p}^{t_s}a_{jl}(t)dt\geq C,
\]
where $C$ is the constant from~\eqref{eq.const-aux}. In particular,
\[
\sum_{j>r}\int_{t_p}^{t_s}f_j(t)dt\geq C(\beta_1-\beta_2).
\]
Applying the Cauchy formula to~\eqref{eq.subsys2} and recalling that $U^{\dagger}$ satisfies~\eqref{eq.column}, one has
\[
\ones_{n-r}^{\top}x^{\dagger}(t_s)\geq\int_{t_p}^{t_s}\ones_{n-r}^{\top}U^{\dagger}(t_s,t)f^{\dagger}(t)dt
\geq \varrho^{\dagger} C(\beta_1-\beta_2).
\]
Hence, for some $j>r$ one has
\[
x_j(t_s)\geq \frac{1}{n-r}\varrho^{\dagger} C(\beta_1-\beta_2)=\beta_2,
\]
which leads to the contradiction.

We have shown that at some time instant $t'\leq t_s$, the solution has at least $r+1$ elements that are $\geq \beta_2$. Using the (backward) induction hypothesis, $\ones^{\top}_nx(t)$ is uniformly positive for $t\ge t'$. Also, $\ones^{\top}_nx(t)\geq r\beta_2$ for $t\in [\xi,t_s]$. This proves the (backward) induction step and finishes the proof of statement (A), which, in turn, implies that~\eqref{eq.column} holds for $n$ agents. Lemma~\ref{lem.tech} is proved.

\subsection{Proof of Theorem~\ref{thm.1}}

Consider a solution of the inequalities~\eqref{eq.ineq2} and the function $\La(t)$ from~\eqref{eq.la}. We know that $\La(t)$ is non-increasing and thus the limit $\La^*=\lim_{t\to\infty}\La(t)\geq -\infty$ exists. If $\La^*=-\infty$, the statement of Theorem~\ref{thm.1} is obvious ($c^*=-\infty$). Henceforth, we assume that $\La^*>-\infty$.

We also introduce the \emph{ordering permutation} of indices $\sigma_1(t),\ldots,\sigma_n(t)$ such that
\[
z_1(t)=x_{\sigma_1(t)}(t)\leq\ldots\leq z_n(t)=x_{\sigma_n(t)}(t).
\]
Notice that, in general, $\sigma_i(t)$ is defined non-uniquely (if $x(t)$ has two or more equal components), however, the permutation $\sigma(t)$ can always be chosen \emph{measurable} and the functions $z_i(t)$ are absolutely continuous on $[0,\infty)$~\cite{TsiTsi:13}.

Using induction on $k=n,n-1,\ldots,1$, we will show that
\be\label{eq.z-consensus}
z_k(t)\xrightarrow[t\to\infty]{}\La^*.
\ee

To prove the induction base $k=n$, notice that by definition
\[
\max_ix_i(\xi)=z_n(\xi)\leq\La(\xi)\forall\xi\geq 0.
\]
The inequality~\eqref{eq.ineq2} and Proposition~\ref{prop.bound} entail that
\[
\dot x_i(t)\leq \alpha_i(t)(\La(\xi)-x_i(t))\quad\forall t\geq\xi\,\forall i.
\]
Using~\eqref{eq.bound}, it can be easily shown that
\[
x_i(t)\leq \theta x_i(\xi)+(1-\theta)\La(\xi)\quad t\in[\xi,\xi+\bar h],
\]
where $\theta=e^{-(n-1)\bar a\bar h}$. Hence, $\La(\xi+h)\leq \theta z_n(\xi)+(1-\theta)\La(\xi)$.
Passing to the limit as $\xi\to\infty$, one shows that $\La_*\leq\varliminf_{\xi\to\infty} z_n(\xi)$. On the other hand, $\varlimsup_{\xi\to\infty}z_n(\xi)\leq\La_*$. The induction base is proved.

Suppose that~\eqref{eq.z-consensus} has been proved for $k=r+1,\ldots,n$. To prove it for $k=r$, it suffices to show that
\be\label{eq.z-consensus1}
\varliminf_{t\to\infty}z_r(t)\geq\La_*.
\ee
Assume, on the contrary, that a sequence $\tau_s\to\infty$ and $\delta>0$ exist such that $z_r(\tau_s)\leq\La(\tau_s)-\delta\,\forall s$. In other words, at $t=\tau_s$ there exists a set of indices $I_s\subset[1:n]$ of cardinality $r$ such that $x_i(t)\leq\La_*-\delta$. Passing to a subsequence, we may assume that $I_s=I$ does not depend on $s$; renumbering the agents, we can suppose that $I=[1:r]$.
Also, without loss of generality, we can assume that $\tau_s=t_{p(s)}$, where $t_p$ is the sequence from either Definition~\ref{def.symm} or Definition~\ref{def.repeated} (recall also that~\eqref{eq.bound} holds). Indeed, if $t_{p(s)-1}<\tau_s=t_{p(s)}$, then
\be\label{eq.aux2}
\begin{gathered}
\dot x_i(t)\leq -\alpha_i(t)[\La(\tau_s)-x_i(t)]\overset{\eqref{eq.bound-alfa}}{\Longrightarrow}\\
x_i(t)\leq \La(\tau_s)-\tilde\delta,\quad \delta_0:=\delta e^{-(n-1)M}\\
\forall i\leq r\,\forall t\in [\tau_s,t_{p(s)}],
\end{gathered}
\ee
For $s$ being sufficiently large, we have $\La(\tau_s)-\La(t_{p(s)})\leq\tilde\delta/2$,
Hence our assumption remains valid, replacing $\tau_s$ by $t_{p(s)}$ and $\delta$ by $\tilde\delta/2$. Next, we consider two cases:

\textbf{Case 1.} Suppose first that condition 2) from Theorem~\ref{thm.1} holds, i.e., $A(\cdot)$ is repeatedly strongly connected. For the index $p(s)$ defined above, let $\tau_s'=t_{p(s)}$.
In view of Proposition~\ref{prop.bound} and~\eqref{eq.bound-alfa}, one shows similar to~\eqref{eq.aux2} that
\be\label{eq.aux2+}
\begin{gathered}
x_i(t)\leq \La(\tau_p)-\delta_0,\quad \delta_0:=\delta e^{-(n-1)M}\\
\forall i\leq r\,\forall t\in [\tau_s,t_{p(s)+1}],
\end{gathered}
\ee
Definition~\ref{def.repeated} ensures the existence of such a pair of indices $j>r$, $i\leq r$ that
\[
M\geq\int\limits_{\tau_s}^{\tau_s'}a_{ji}(t)dt\geq\ve.
\]
By noticing that
\[
\dot x_j(t)\leq -\alpha_j(t)x_j(t)+\La(\tau_s)\alpha_j(t)-a_{ji}(t)\delta_0
\]
for $t\in[\tau_s,t_{p(s)+1}]$, one shows that $x_j(\tau_s')\leq \La(\tau_s)-\delta'$, where
$\delta'=\delta_0\ve e^{-M}$. We have shown that at time $t=\tau_s'$
the vector $x(t)$ has at least $r+1$ components that are $\leq \La(\tau_s)-\delta'$, that is,
$z_{r+1}(\tau_s')\leq\La(\tau_s)-\delta'$. Passing to the limit as $s\to\infty$, we arrive at a contradiction with the induction hypothesis. Hence,~\eqref{eq.z-consensus1} is valid.

\textbf{Case 2.} The case of type-symmetric matrix is less trivial. In this case, we employ Lemma~\ref{lem.tech}, and the structure of the proof is similar. We introduce two delay inequalities of smaller dimensions
\begin{gather}
\dot x_i\leq\sum\limits_{j\leq r}a_{ij}(t)(\hat x_j^i(t)-x_i(t))+f_i(t),\,i=1,\ldots,r;\label{eq.subsys1+}\\
\dot x_i\leq\sum\limits_{j> r}a_{ij}(t)(\hat x_j^i(t)-x_i(t))+f_i(t),\,r<i\leq n.\label{eq.subsys2+}
\end{gather}
and notice that the subvectors $x^+(t)=(x_i(t))_{i\leq r}$ and $x^{\dagger}(t)=(x_i(t))_{i>r}$ satisfy, respectively, to~\eqref{eq.subsys1+} and~\eqref{eq.subsys2+} with $f_i$ defined as follows:
\be\label{eq.aux-f}
\begin{gathered}
f_i(t)=
\begin{cases}
\sum_{j>r}a_{ij}(t)[\hat x^i_j(t)-x_i(t)],\quad i\leq r\\
\sum_{j\leq r}a_{ij}(t)[\hat x^i_j(t)-x_i(t)],\quad i>r.
\end{cases}
\end{gathered}
\ee

Arguments similar to~\eqref{eq.aux2} allow to prove the following fact.
Consider the ``undisturbed'' ($f_i\equiv 0$) system of equations associated to~\eqref{eq.subsys1+}
\begin{gather}
\dot x_i=\sum\limits_{j\leq r}a_{ij}(t)(\hat x_j^i(t)-x_i(t)),\,i=1,\ldots,r;\label{eq.subsys1}
\end{gather}
and consider its solution defined for $t\geq \tau_s$ with such an initial condition that
\be\label{eq.aux3}
x_i(\tau_s)\leq\La(\tau_s)-\delta,\quad x_i(t)\leq\La(\tau_s)\,\forall t<\tau_s.
\ee
Then $x_i(t)\leq \La(\tau_s)-\delta_1\,\forall t\geq \tau_s$, $\delta_1=\delta e^{-(n-1)\bar a\bar h}$. Indeed, this inequality holds for $t\in [\tau_s,\tau_s+\bar h]$, where for $t\geq\bar h+\tau_s$ it is immediate from Proposition~\ref{prop.bound} applied to~\eqref{eq.subsys1}. Using Remark~\ref{rem.positive}, one shows that any solution of \emph{inequality}~\eqref{eq.subsys1+} with initial condition satisfy~\eqref{eq.aux3} admits the following estimate
\be\label{eq.aux4a}
x^+(t)\leq (\La(\tau_s)-\delta_1)\ones_r+\int_{\tau_s}^{t}U^+(t,\xi)f^+(\xi)d\xi.
\ee
Applying Remark~\ref{rem.positive} to the equation~\eqref{eq.subsys2+} and associated inequality~\eqref{eq.subsys2+}, one shows that
\be\label{eq.aux4b}
x^{\dagger}(t)\leq \La(\tau_s)\ones_{n-r}+\int_{\tau_s}^{t}U^{\dagger}(t,\xi)f^{\dagger}(\xi)d\xi.
\ee

Let $t_*\geq\tau_s$ be the first instant such that $x_i(t_*)=\La(\tau_s)-\delta/2$ for some $i\leq r$; we put $t_*=\infty$ if such an instant does not exist. If $t_*<\infty$, let $\tilde\tau_s$ be the instant $t_q$ such that $t_{q-1}\leq t_*\leq t_q$, otherwise, $\tilde\tau_s=\infty$. Then, similar to~\eqref{eq.aux2}, for all $i\leq r$ and $t\in [\tau_s,t_q]$ one has
\[
x_i(t)\leq \La(\tau_s)-\delta_2,\quad\delta_2:=\frac{\delta}{2}e^{-(n-1)M}.
\]
We are going to prove that for some $\ell>r$ and some $t'\in [\tau_s,\tilde\tau_s]$ we have $x_{\ell}(t')\leq \La(\tau_s)-\delta_3$, where
\be\label{eq.const-aux+}
\begin{gathered}
\delta_3:=\frac{\varrho^{\dagger}C_1\delta_2}{\varrho^{\dagger}C+n-r},\
C_1:=\frac{\delta}{2Kr(n-r)},
\end{gathered}
\ee
and $\varrho^{\dagger}$ is the constant from Lemma~\ref{lem.tech}, corresponding to $U^{\dagger}$.

Assuming the contrary, we have $\hat x_j^i(t)-x_i(t)\geq (\delta_2-\delta_3)$ whenever $i\leq r<j$ and $t\in [\tau_s,\tilde\tau_s]$. In the case where $\tilde\tau_s=t_*=\infty$, we thus have $\sum_{j>r}\int_{\tau_s}^{\infty}f_j(t)dt=-\infty$ since the persistent graph is connected and thus contains at least one arc $(i,j)$, where $i\leq r<j$. Using~\eqref{eq.aux4b} and Lemma~\ref{lem.tech}, we have
\[
\ones_{n-r}^{\top}x^{\dagger}(t)\xrightarrow[t\to\infty]{}-\infty,
\]
arriving thus at a contradiction. In the case where $t_*<\infty$, we use~\eqref{eq.aux4a} and the assumption $x_i(t_*)=\La(\tau_s)-\delta/2$ to derive, similar to~\eqref{eq.auxaux}, that
\be\label{eq.auxaux}
\int_{t_{p(s)}}^{t_*}a_{lj}(t)[\hat x^l_j(t)-x_l(t)]dt\geq c_1:=\frac{\delta}{2r(n-r)}.
\ee
for some $l\leq r$ and $j>r$. Therefore,
\[
\int_{t_{p(s)}}^{t_q}a_{lj}(t)dt\geq\int_{t_{p(s)}}^{t_*}a_{lj}(t)dt\geq c\overset{\eqref{eq.type-symm-non}}{\Longrightarrow}
\int_{t_p}^{t_s}a_{jl}(t)dt\geq C_1,
\]
where $C_1$ is the constant from~\eqref{eq.const-aux+}. In particular,
\[
\sum_{j>r}\int_{t_p}^{t_s}f_j(t)dt\leq -C_1(\delta_2-\delta_3).
\]
Using~\eqref{eq.aux4b} and recalling that $U^{\dagger}$ satisfies~\eqref{eq.column}, one has
\[
\begin{gathered}
\ones_{n-r}^{\top}x^{\dagger}(t_q)\leq (n-r)\La(\tau_s)+
\int_{t_{p(s)}}^{t_q}\ones_{n-r}^{\top}U^{\dagger}(t_q,t)f^{\dagger}(t)dt\\
\leq (n-r)\La(\tau_s)-\varrho^{\dagger} C_1(\delta_2-\delta_3).
\end{gathered}
\]
Hence, for some $j>r$ one has
\[
x_j(t_q)\leq \La(\tau_s)-\frac{1}{n-r}\varrho^{\dagger} C_1(\delta_2-\delta_3)=\La(\tau_s)-\delta_3,
\]
which leads to the contradiction. Hence, for each $s$ there exists $t'_s>\tau_s$ such that the vector $x(t'_s)$ has at least $r+1$ components less than $\La(\tau_s)-\delta_3$, so $z_{r+1}(t'_s)\leq \La(\tau_s)-\delta_3$. Passing to the limit as $s\to\infty$, one arrives at the contradiction with the induction hypothesis. This finishes the proof of~\eqref{eq.z-consensus1} in the second case.

The property~\eqref{eq.delta0} can be proved similarly to Remark~4 in~\cite{ProCao:2017}.
Theorem~\ref{thm.1} is proved.

\bibliographystyle{IEEETran}
\bibliography{consensus,social}

\end{document}